\documentclass[10pt]{raa}            
\usepackage[T1]{fontenc}
\usepackage{ae,aecompl}
\usepackage{graphicx,times}             
\begin{document}

   \title{Numerically investigating  the peculiar periphery of a supernova remnant in the medium with a density gradient: the case of RCW 103
}

   \volnopage{Vol.0 (20xx) No.0, 000--000}      
   \setcounter{page}{1}          

   \author{Chun-Yan\ Lu
      \inst{1}
      \and Jing-Wen\ Yan
      \inst{1}
   \and Lu\ Wen
      \inst{1}
   \and Jun\ Fang
      \inst{1}
   }

   \institute{Department of Astronomy, Key Laboratory of Astroparticle Physics of Yunnan Province, Yunnan University, Kunming 650091, China; {\it fangjun@ynu.edu.cn}
   }

   \date{xxx}

\abstract{
The young shell-type supernova remnant RCW 103 has  peculiar properties in the X-ray morphology obtained with Chandra. The southeastern shell is more brighter in the X-rays, and the curved border of the shell in this region is more flatten than the other part. We investigate the formation of the peculiar periphery of the supernova remnant RCW 103 using 3D hydrodynamical simulation. Assuming that the supernova ejecta has been evolved in the medium with a density gradient, the detected shape of the periphery can be generally reproduced. For RCW 103, with the ejecta mass of $3.0~M_{\odot}$, the density of the background material of $2.0~\mathrm{cm}^{-3}$, and a gradient of $3.3 - 4.0 ~\mathrm{cm}^{-3} \mathrm{pc}^{-1}$, the X-ray periphery can be generally reproduced.  The simulation turned out that the asymmetry of the SNR  RCW 103 is mainly due to the inhomogeneous medium with a density gradient.
}

   \authorrunning{Lu et al. }            
   \titlerunning{Numerically investigating  the peculiar periphery of RCW 103}  

   \maketitle

%
%
\section{Introduction}
\label{intro}

The distribution of the medium around a supernova remnant (SNR) gives impact on the morphology of the remnant, which is partly the reason for the diversity of the morphologies of SNRs. RCW 103 is a typical SNR which has an asymmetrical morphology and a noncircular periphery in the X-rays. It has a radius of 9~pc with a distance of 3.3~kpc, which is derived from the HI absorption measurements (Caswell et al~\cite{CR75}; Caswell et al~\cite{CH80}). The age is estimated to be $\sim2000$ yr through the average optical expansion rate of the outer edge about 1100 $\mathrm {km / s}$~(Carter \& Dickel et al.~\cite{CD97}). There is a strong X-ray emission region in the southeast of the remnant in the morphologies obtained with Chandra and XMM-Newton (Caswell et al.~\cite{CH80}; Dickel et al.~\cite{DG96}; Rho et al.~\cite{RR01}; Pinheiro Gon$\c{c}$alves et al.~\cite{PN11}; Braun et al.~\cite{BS19}). We note that the remnant has a central low-density area from southwest to northeast (`bar'). Bear \& Soker (\cite{BS18}) suggest that two opposite jets that the newly born neutron star launched during the explosion shaped this low density bar, or more generally the barrel-shape morphology (Akashi et al. \cite{AB18}). RCW 103 has converted from double-shock stage (Chevalier et al.~\cite{C82}) dominated by the ejected material, to the point-blast stage (Sedov et al.~\cite{S59}), which relied on the swept-up circumambient material. The observations about $H_{2}$ of 2.121 $\mu m$ line or others suggested that RCW 103 might interact with molecular material in the south limb (Oliva et al.~\cite{OM90}; Rho et al.~\cite{RR01}; Paron et al.~\cite{PR06}).

Numerical simulations have been used to study morphologies and evolutions of SNR extensively. Toledo-Roy et al. (~\cite{TV14}) explained the morphology of SNR G352.7-0.1 through the blowout scenario. In the model, the supernova explosion occurred inside and near the border of a spherical cloud, with the aim of the remnant blowout into the lower-density ambient medium. They found that the quasi-perpendicular and quasi-parallel mechanisms played an important part in particle acceleration efficiency and produced multiple ring-like or arc-like structures. Moranchel-Basurto et al (~\cite{MV17}). considered that the SNR G296.5+10 had expanded into ambient medium by means of 3D MHD simulations, where the magnetic field was a superposition of magnetic field of the progenitor star and the galactic magnetic field. They concluded that quasi-parallel acceleration mechanisms in the shock front can explain the morphology and the problem of different rotation measure in the east and west. Wu \& Zhang (\cite{WZ19}) studied the evolution of SNRs in a strong magnetic field. They considered the existence of strong magnetic fields such as those inside galaxies. In the simulation, when the magnetic field of 1~mG, most of the ejecta will propagate parallel to the magnetic field. Fang et al. (~\cite{FY20}) used 3D HD simulations to investigate the morphology of SNR SN 1006, taking into account of supernova ejecta evolved in the surrounding medium with a discontinuous density. The interaction between SNR and complex interstellar media can produce asymmetric structures. For example, Kosti$\acute{c}$ et al. (\cite{KK19}) studied the large-scale interaction between supernova remnants and interstellar gas clouds. They used hydrodynamics to simulate the interstellar clouds with fractal density structure, which affected the expansion and shock properties of the remnants. Zhang $\&$ Chevalier (\cite{ZC19}) studied the interaction between SNR and a turbulent molecular cloud medium, they found that in turbulent media with high Mach number, the SNR has lower interior temperature, lower radial momentum, and dimmer X-ray emission compared to one in a less turbulent medium with the same mean density. The interaction between the precursor stars of supersonic motion and the interstellar medium can produce interesting results. This model can explain the strange shape of the Galactic SNR VRO 42.05.01.(Chiotellis et al. \cite{CB19}; Derlopa et al. \cite{DB19}).

In this paper, we simulate the morphology of the SNR RCW 103 with the assumption that it interacts the ambient medium with a density gradient using 3D HD simulations. In section 2, we interpret the  model and numerical set-up. In section 3, we give the results of the simulation. In section 4, the discussion and some conclusions are given.
\section{THE MODEL AND NUMERICAL SET-UP}
\label{simuset}
\subsection{Numerical method}
\label{HDmodel}
The dynamical evolution plays a crucial part to form peculiar morphological SNRs. In the simulations, we adopt the PLUTO code (Mignone et al.~\cite{MB07}, ~\cite{MZ12}) and 3D Cartesian coordinate system, ignoring the radiative cooling, the thermal conduction, and the influence of the magnetic field in the dynamics. The equations of mass, momentum, energy conservation are as follows,
\begin{eqnarray}
\frac{\partial\rho}{\partial t} + \nabla\cdot(\rho \textbf{v}) & = & 0\; , \\
\frac{\partial \rho {\bf v}}{\partial t} + \nabla \cdot ( \rho {\bf
    vv} )   + \nabla{P} & = & 0\; , \\
\frac{\partial E}{\partial t} + \nabla \cdot (E+P){\bf v} ) & = & 0 ,
\end{eqnarray}
where $\rho$ is the matter density,   $P$ is the gas pressure, $\mathbf{v}$ is the gas velocity, and
\begin{equation}
E = \frac{P}{\gamma - 1}+\frac{1}{2}\rho v^2 \;,
\nonumber
\end{equation}
is the total energy density. The adiabatic index for the nonrelativistic gas is adopt to be $\gamma=3/5$, $\rho = \mu m_{H} n_{H}$ is the mass density, $m_{H}$ is the mass of the hydrogen atom, $n_{H}$ is the hydrogen number density, and $\mu = 1.4$ is the mean atomic mass assuming a $10 : 1$ $\mathrm{H}:\mathrm{He}$ ratio.

\subsection{Set-up of the ejecta}
\label{setup}
$E_{ej}$ and $M_{ej}$ are the kinetic energy and the mass of the ejecta, respectively, and the index $s$ is adopted to be $9$ for the core-collapse supernova. The ejecta has a homogeneous density within a radius of $r_{c}$ and a power-law distribution on $r$ with $r>r_{c}$, i.e.,
\begin{equation}
\rho_{\rm ej}(r) = \left\{
  \begin{array}{cc}
    \rho_{\mathrm c},& \mathrm{if~} r < r_{\mathrm c}\ \\
    \rho_{\mathrm 0}(r/R_{\mathrm ej})^{-s}, & \mathrm{if~} r_{\mathrm c} < r <  R_{\mathrm{ej}}\
    \label{rho_c}
   \end{array}
\right.
\end{equation}
and
\begin{equation}
r_{\mathrm c}= R_{\mathrm{ej}} \left[1- \frac{\eta(3-s)M_{\mathrm{ej}}}{4\pi\rho_{0}R_{\rm ej}^3} \right]^{\frac{1}{3-s}} ,
\end{equation}
where $\rho_{0}$ is the density at $r = R_{ej}$. $R_{ej}$=1.5~pc is the incipient radius of the ejecta, and $\eta=3/7$ is the mass ratio of the outer layer of the ejecta to the entire ejecta (Jun \& Norman~\cite{JN96}). We acquire the uniform density through (Jun \& Norman~\cite{JN96})
\begin{equation}
\rho_{\mathrm c}= \frac{3(1-\eta)M_{\mathrm{ej}}}{4\pi r_{\rm c}^3} ,
\end{equation}
and the velocity of the ejecta at the outer edge $v_{0}$ is
\begin{equation}
v_{0} = (E_{\rm ej})^{1/2}\left \{\frac{2\pi \rho_{\rm c}r_{\rm c}^5}{5R_{\rm ej}^2}
+ \frac{2\pi \rho_{R}R_{\rm ej}^3\left[1-(R_{\rm ej}/r_{\rm c})^{s-5}\right]}{5-s}\right \}^{-1/2}\;.
\end{equation}

\subsection{Modelling the density-gradient}
To investigate the dynamical evolution of the SNR RCW 103 to reproduce the peculiar morphology, and we assuming that the ejecta has evolved in the ambient medium with density gradient, and the number density of the ambient medium is assumed to be (Yan et al. \cite{YL20})
\begin{equation}
n(\mathbf{r}) = \left\{
  \begin{array}{cc}
    n_{0}  + k\widehat{\xi}\cdot \mathbf{r},  & \mathrm{if~} \widehat{\xi}\cdot \mathbf{r}>0 \\
    n_{0}, & otherwise \
    \label{nr}
   \end{array}
\right.
\end{equation}
Where $\mathbf{\widehat{\xi}} = \sin\theta \cos\phi\hat{e}_x + \sin\theta \sin\phi\hat{e}_y + \cos\theta\hat{e}_z$ is the unit vector of the transformation between Cartesian coordinate system and spherical coordinate system, and the density gradient medium is distributed along this direction,  $k$ is the magnitude of the density gradient, and $n_{0}$ is the total particle number density of the ambient medium.

\section{Results}
\label{resu}

\begin{figure}[tbh]
\begin{center}
\includegraphics[width=0.6\textwidth]{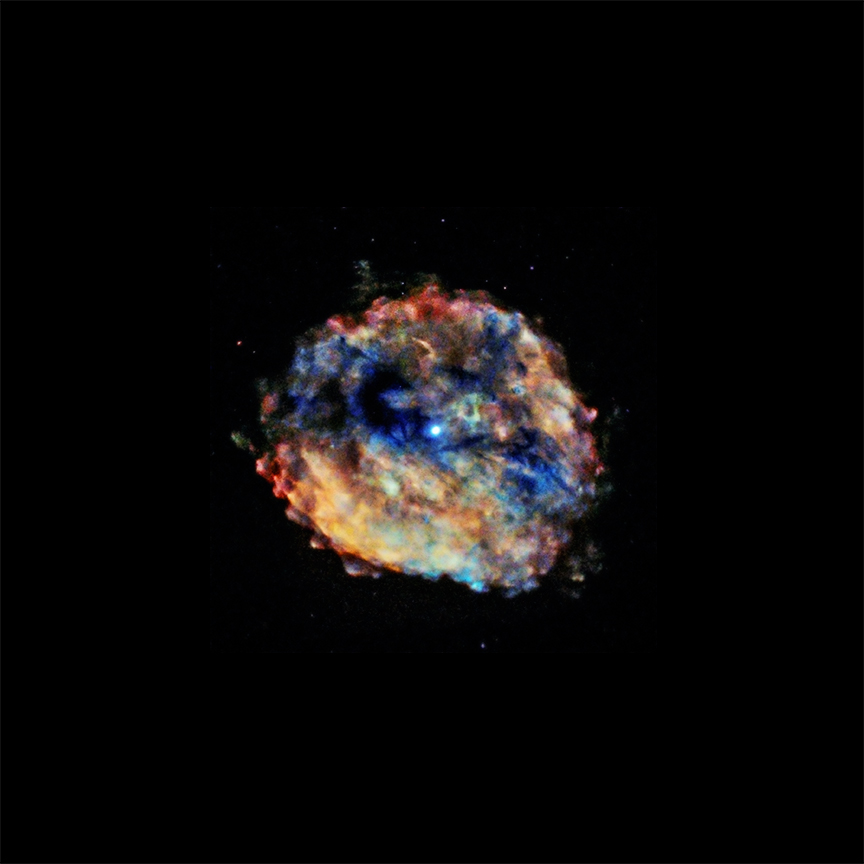}
\end{center}
\caption{The observed X-ray morphology of SNR RCW 103 with Chandra, lowest energy (red),the medium band (green) and X-ray (blue). (This image was downloaded from https://chandra.harvard.edu/photo/2016/rcw103/.)}
\label{fig:chandra}
\end{figure}

As illustrated in Fig.\ref{fig:chandra}, the image of RCW 103 obtained by Chandra indicates that the shell in the southeast is more flattened and brighter than the northwest and the periphery deviates from circular (Frank et al 2015). To reproduce this peculiar periphery, we assume the ejecta has evolved in the medium with a density gradient.
The initial conditions are set in a cartesian coordinate system which extends from $-12~\mathrm{pc}$ to $12~\mathrm{pc}$ in each of the $x$, $y$ and $z$ directions. The density gradient is along $\mathbf{\widehat{\xi}} = \sin\theta \cos\phi\hat{e}_x + \sin\theta \sin\phi\hat{e}_y + \cos\theta\hat{e}_z$. Moreover, the total mass of the ejecta is $3.0~M_{\odot}$, and the explosion of kinetic energy is $1.0 \times 10^{51}~\mathrm{erg}$. Then the maximum velocity of the material at the boundary of the ejecta is $v_{0}=2 \times 10^{9}~\mathrm{cm~s}^{-1}$. The SNR has evolved in the medium with $n_{0} = 2.0~\mathrm{cm}^{-3}$. The other parameters adopted in the simulation are listed in Table \ref{tab:0} for the six different models with different density gradient ratio and direction.

\begin{table}
\begin{center}
\caption[]{Parameters for the SNR RCW 103 with $M_{\mathrm{ej}}=3.0~M_{\odot}$, $E_{ej} = 10^{51}~\mathrm{erg}$, $R_{\mathrm{ej}}=1.5~\mathrm{pc}$, $n_{0}=2.0~\mathrm{cm}^{-3}$, $T = 10^{4}\mathrm{K}$, $\eta=3/7$. }\label{Tab:para1}
 \begin{tabular}{ccccccc}
   \hline\noalign{\smallskip}
   Parameters    &  Model A  &  Model B  &  Model C  &   Model D  &   Model E   &   Model F  \\
   \hline\noalign{\smallskip}
   $\theta(^{\circ})$   &   $200$  &    $200$    &     $200$     &    $200$    &     $200$      &    $190$  \\
   $\phi(^{\circ})$   &   $90$  &    $90$    &     $90$     &    $90$    &     $45$      &    $90$  \\
   $k(cm^{-3} pc^{-1})$      &     0.5       &      1.3      &      2.0      &     2.5      &       1.3      &      1.3      \\
   \noalign{\smallskip}\hline
\label{tab:0}
\end{tabular}
\end{center}
\end{table}

\begin{figure}[tbh]
\begin{center}
\includegraphics[width=1.0\textwidth]{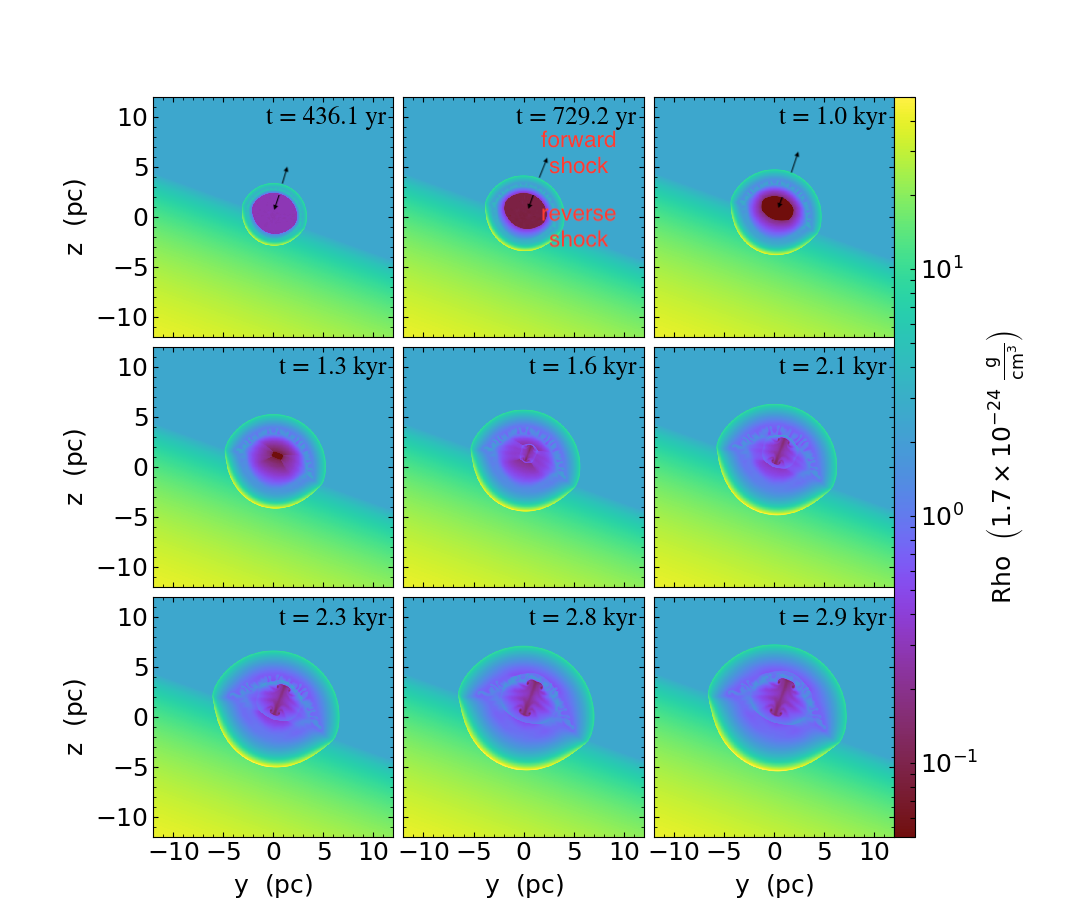}
\end{center}
\caption{The slices of density with different times at $x=0$ for the Model B. The detail of the parameters is indicated in Tabel \ref{tab:0}.}
\label{fig:prrho}
\end{figure}

\begin{figure}[tbh]
\begin{center}
\includegraphics[width=1.0\textwidth]{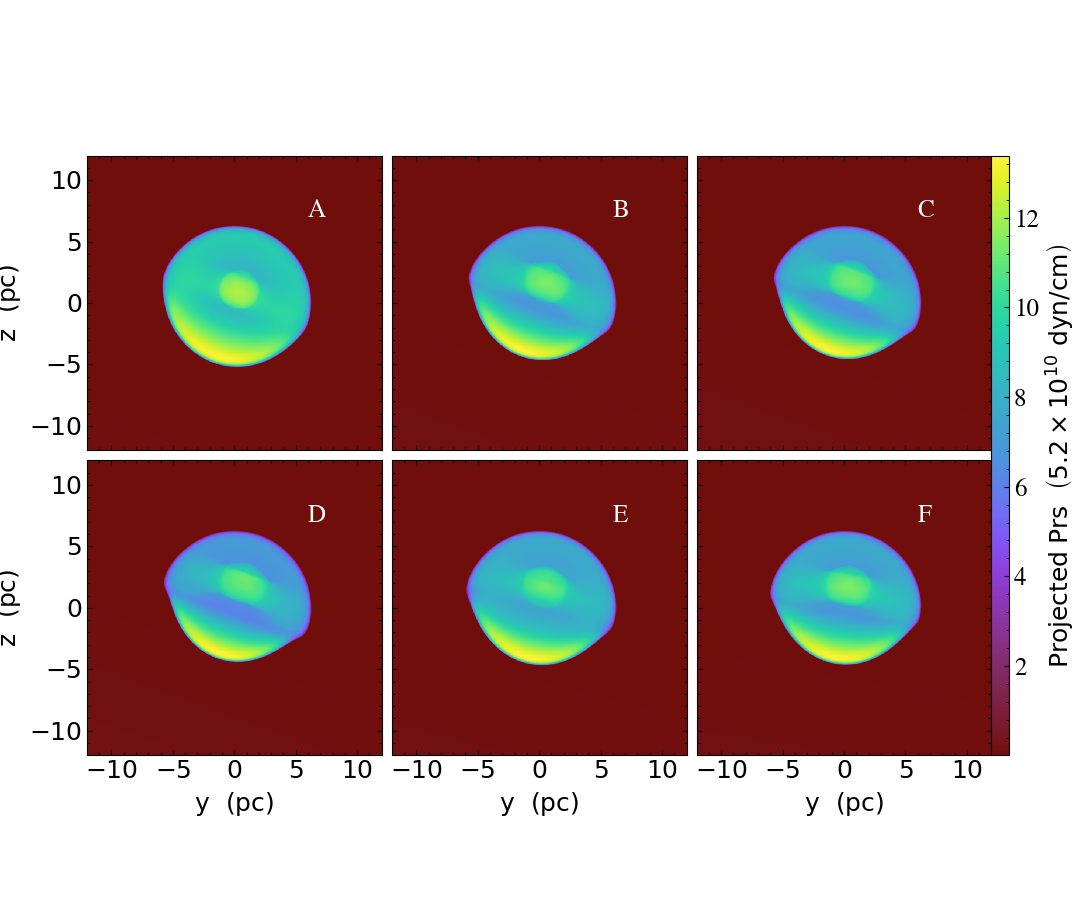}
\end{center}
\caption{The projected pressure along $x$ for the six models with the parameters listed in Tabel \ref{tab:0}.}
\label{fig:prpro}
\end{figure}

Fig.\ref{fig:prrho} shows the evolution of the density in the plane $x=0$ for the Model B, in which a density gradient of the ambient medium directed to the southeast. The ejecta produces a forward shock wave due to the supersonic motion, and it compresses and thermalizes the ambient matter constantly.  The thermalized medium also produce a reverse shock that constantly compresses the matter. Moreover, the finger-like and mushroom-like structures are produced around the contact discontinuity location due to Rayleigh-Taylor instabilities. In the southeast, the forward shock is encountered with a higher-density medium compared with the northwest, which results in the velocity of the forward shock in the southeast is smaller than than the others. Reverse shocks are reflected when they come across centrally and the reflected reverse shocks make the ejecta thermalized. Our simulations give a `bar' of low density (with the mushroom shape on both sides) that is perpendicular to the low density bar in the X-ray image (Fig.\ref{fig:chandra}). The observed low density region results from another process. For example, Bear \& Soker (\cite{BS18}) suggest that two opposite jets that that were launched at explosion shaped this low density region.

In Fig.\ref{fig:prpro}, we show the projected pressure along the $x$ direction at an age of $2.1$~kyr for the six models. The two shock structure is formed with a forward shock and a reverse shock for all the six model. The influence of $k$ on the resulting periphery of the remnant can be seen in the panels for the Model A to D with $\theta = 200^{\circ}$, $\phi =200^{\circ}$ and $k = 0.5, 1.3, 2.0$ and $2.5~cm^{-3}pc^{-1}$, respectively. With a larger density gradient (a larger $k$), the extent to which the remnant is squeezed by the medium in the southeast was more significant. In the model A, the contour of the remnant is approximately circular, and the ejecta in the southeast has a larger pressure due to the denser medium. In model D, the outline of the remnant deviates significantly from circular, the radius in the southeast is smaller than in the northwest.
The density gradient of the Model B is the same as the Model E, but the directions for the two model is different with $\phi=90^{\circ}$ and $45^{\circ}$, respectively. The change of angle $\phi$  corresponds to a rotation in the angle of line of sight. The pole of the maximum projected pressure directs along the direction of the density gradient (see the panels for Model E and Model B).
As illustrated in the panel for the Model B in Fig.\ref{fig:prpro}, the pressure in the southeast shell is higher than others, and the single bright structure  is consistent with the bright feature in the southeast in the X-ray image.

\section{Summary and discussion}
\label{sumdis}
In this paper, we investigate the reason for the formation of the peculiar periphery as indicated in the  X-ray image for the SNR RCW 103 based on 3D HD simulation. The shell is brighter and the boundary is more flatten in the southeast than in the other part. We assume the SNR evolves in a complex environment with a density gradient in a specific direction to reproduce the peculiar periphery of the remnant. The structure of the SNR is asymmetric due to the interaction of the forward shock with the inhomogeneous medium. Based on the images observed by Chandra, the main shell of SNR RCW 103 has two hemispheres with different radius in the northwest and southeast directions, and the radius ratio is about 1:1.08 (the radius ratio is divided by the distance from the center brightest point to the northwest and southeast edges respectively), In the southeast, the periphery in the X-ray image can be generally reproduced with a density gradient of $3.3 - 4.0~ \mathrm{cm}^{-3}\mathrm{pc}^{-1}$. Our simulations do not reproduce the low density bar seen in observations.   It is possible that jets at explosion formed this structure (Bear \& Soker \cite{BS18}).

We just pay attention to the peculiar periphery of the SNR RCW 103 as illustrated by the resulting projected pressure, and the morphology of emission is not derived because otherwise more assumption on the distribution of high-energy particle and the magnetic field are needed.

\begin{acknowledgements}
JF is partially supported  by the
Natural Science Foundation of China (NSFC) through grant 11873042, the National Key R\&D Program of China under grant No.2018YFA0404204,
the Yunnan Applied Basic Research
Projects under (2018FY001(-003)), the Candidate Talents Training Fund of Yunnan Province (2017HB003)
and the Program for Excellent Young Talents, Yunnan University (WX069051, 2017YDYQ01).
\end{acknowledgements}

\label{lastpage}

\end{document}